\documentclass{WileyMSP-template}

\usepackage{geometry}
\usepackage{breakurl}
\usepackage{amsfonts}
\usepackage{amsmath}
\usepackage{amssymb}
\usepackage{mathrsfs}
\usepackage{epsfig}
\usepackage{graphicx}
\usepackage{bm}
\usepackage{color,xcolor}
\usepackage[numbers,sort&compress]{natbib}

\begin{document}


\title{{Exceptional Photon Blockade: Engineering Photon Blockade with Chiral Exceptional Points}}

\maketitle


\author{Ran Huang}
\author{{\c{S}}. K. {\"O}zdemir*}
\author{Jie-Qiao Liao}
\author{Fabrizio Minganti}
\author{Le-Man Kuang}
\author{Franco Nori*}
\author{Hui Jing*}



\begin{affiliations}
Dr. Ran Huang, Prof. Jie-Qiao Liao, Prof. Le-Man Kuang, Prof. Hui Jing\\
Key Laboratory of Low-Dimensional Quantum Structures and Quantum Control of Ministry of Education, Department of Physics and Synergetic Innovation Center for Quantum  Effects and Applications, Hunan Normal University, Changsha 410081, China\\
Email Address: jinghui73@foxmail.com

Prof. {\c{S}}ahin K. {\"O}zdemir\\
Department of Engineering Science and Mechanics, and Materials Research Institute (MRI), The Pennsylvania State University, University Park, PA 16802, USA\\
Email Address: sko9@psu.edu

Dr. Ran Huang, Dr. Fabrizio Minganti, Dr. Franco Nori\\
Theoretical Quantum Physics Laboratory, RIKEN Cluster for Pioneering Research, Wako-shi, Saitama 351-0198, Japan\\
Email Address: fnori@riken.jp

Dr. Franco Nori\\
Physics Department, The University of Michigan, Ann Arbor, Michigan 48109-1040, USA

\end{affiliations}



\begin{abstract}

Non-Hermitian spectral degeneracies, known as exceptional points (EPs), feature simultaneous coalescence of both eigenvalues and the associated eigenstates of a system. A host of intriguing EP effects and their applications have been revealed in the classical realm, such as loss-induced lasing, single-mode laser, and EP-enhanced sensing. Here we show that a purely quantum effect, known as single-photon blockade, emerges in a Kerr microring resonator due to EP-induced asymmetric coupling between the optical modes and the nonlinearity-induced anharmonic energy-level spacing. A striking feature of this photon blockade is that it emerges at two-photon resonance which in Hermitian systems will only lead to photon-induced tunneling but not to photon blockade. By tuning the system towards to or away from an EP, one can control quantum correlations, implying the potential use of our system for frequency tunable single-photon generation and an antibunching-to-bunching light switch. Our work sheds new light on EP-engineered purely quantum effects, providing unique opportunities for making and utilizing various single-photon quantum EP devices.

\end{abstract}


\section{Introduction}
{\quad}Non-Hermitian systems exhibit a vast range of counterintuitive phenomena and substantially different effects, such as loss-induced lasing~\cite{peng2014loss} and chiral perfect absorption~\cite{Soleymani2022Chiral}, with no correspondence in their Hermitian counterparts, due to to their spectral degeneracies, known as exceptional points (EPs)~\cite{Bender2007making,Sahin2019PT,Ganainy2019the}. Different from Hermitian spectral degeneracies, referred to as diabolic points (DPs) which feature degenerate eigenvalues with orthogonal eigenstates, EPs feature degeneracy in both the eigenvalues and their associated eigenstates, leading to a reduction in the system's dimensionality and a very skewed vector space. Therefore, when the eigenenergy maps are reconstructed by steering the system's parameters, a Hermitian system exhibits a double-cone topology which locates the DP at the apex where the cones touch each other, while a non-Hermitian system exhibits a complex-square-root topology of two intersecting Riemann sheets with a branch point singularity at the EP.

{\quad}In recent years, peculiar features of non-Hermitian systems have been utilized as resources to construct unconventional devices to control the flow of light and its interaction with matter~\cite{Feng2011nonreciprocal,Peng2014parity,Peng2016chiral,Zhong2019Controlling,ruter2010observation,jing2014PT,Jing2015Optomechanically,Lv2018Optomechanically,Jing2017High,Zhang2018LIT,Lu2021Exceptional,xu2016topological,bender2013observation,bender2013observationpt,zhu2014pt,popa2014non,fleury2015invisible,shi2016accessing,ding2018anisotropicEP,qiu2019antipt,Zhang2020Breaking,zhang2022AntiPTsymmetric}. EP-enabled classical devices, such as single-mode lasers~\cite{Feng2014single,Hodaei2014parity}, wireless power transfer~\cite{assawaworrarit2017robust}, sensors~\cite{chen2017exceptional,dong2019sensitive,Zhong2019Sensing,Lai2019observation,Hokmabadi2019non,chen2018generalized}, and topological devices~\cite{Gao2015Observation,Zhang2020Tunable}, have been demonstrated. Very recently, EPs have been studied also in purely quantum systems~\cite{Dujf2019quPT,klauck2019observation,Naghiloo2019quantum,Cao2020Reservoir}, inspiring a search for EP-tuned quantum effects and their unique applications~\cite{yuan2020Steady,perina2019Nonclassical}.


\begin{figure}[t!]
  \includegraphics[width=\linewidth]{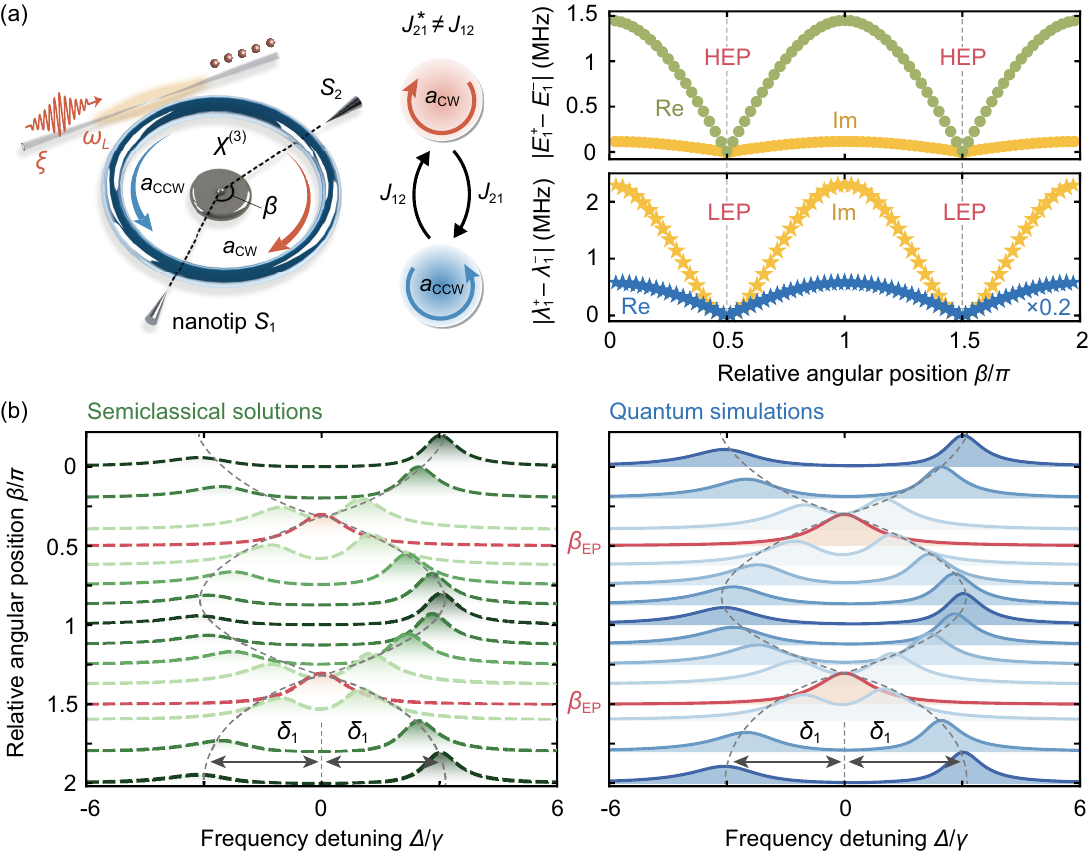}
  \centering
  \caption{{Periodic exceptional points (EPs) in a non-Hermitian system.} {(a)} EPs in a resonator with strong Kerr nonlinearity $\chi^{(3)}$ driven by a laser field with frequency $\omega_{L}$ and amplitude $\xi$. Both Hamiltonian exceptional points (HEPs) and Liouvillian exceptional points (LEPs) emerge at $\beta/\pi$$\ =\ $$\{0.5,1.5\}$. Here, Re (Im) denotes the real (imaginary) part of the eigenvalues. {(b)} Normalized cavity excitation spectrum $S_{11}(\Delta)$ obtained using the semiclassical (left) and quantum (right) methods. Here, $\delta_1$$\ =\ $$\sqrt{J_{12}J_{21}}$. For experimentally accessible parameter values, see the main text. }
  \label{Fig1}
\end{figure}

{\quad}In this work, we show that a truly \emph{quantum} effect, i.e., photon blockade (PB), can be well tuned by the interplay of EPs and nonlinearity of the system. As a manifestation of the quantization of light, PB refers to the process in which the absorption of one photon blocks the absorption of subsequent ones~\cite{imamoglu1997strongly}. PB plays key roles in building single-photon devices and creating non-classical correlations for applications in quantum engineering, as already studied in a wide range of physical systems~\cite{Birnbaum2005,Greentree2006Quantum,Faraon2008,reinhard2012strongly,muller2015coherent,Adam2014State,huang2018nrpb,Lang2011,Liu14pb,rabl2011photon,liao2013photon,wang2015tunable,You2020Reconfigurable}.
So far, PB has been classified into three groups: (i) Conventional photon blockade (CPB), which emerges when optical interactions detune the states with high number of photons, leaving the few-photon manifold unaffected~\cite{Birnbaum2005,Faraon2008,shamailov2010multi,adam2013two,hamsen2017two,Kowalewska2019Two,Delteil2019towards}; (ii) unconventional photon blockade (UPB), which occurs due to destructive interferences among different transition pathways~\cite{Liew2010single,bamba2011origin,Flayac2017unconventional,Snijders18,Vaneph18,li2019nrupb}; and (iii) PB induced by the truncation of Hilbert space (TPB: truncation-induced PB), as observed in quantum linear scissors~\cite{Sahin2001Quantum,Lvovsky2002Quantum,sahin2002Pulse}.
For all these cases, single-photon blockade cannot occur under two-photon or multi-photon resonance conditions~\cite{Faraon2008,Kowalewska2019Two}, because the involvement of two or more photons indicate that the absorption of the first photon favors also that of subsequent photons, i.e., resulting in bunched light.

{\quad}Here, we show that, contrary to general belief, single-photon blockade under two-photon resonance can take place at chiral EPs in a nonlinear Kerr resonator. {The basic principle underlying this counterintuitive effect is the interplay of the EP-induced chirality and the nonlinearity-induced anharmonic energy-level spacing}. Complete localization of a single photon in either the clockwise (CW) or the counterclockwise (CCW) mode occurs when the system is operated at EPs. Namely, the asymmetric coupling between the optical modes of a microring resonator creates periodic EPs~\cite{chen2017exceptional,Peng2016chiral} that impose a strong spatial chirality on the coalesced modes~\cite{Peng2016chiral}. Exactly at the EPs, the modes become fully chiral, such that the modes propagate in only one direction. Thus, a single photon is localized in the CW or the CCW mode when the system is operated at the EPs, leading to a switchable chirality. We note that this process does not rely on introducing any gain into the system~\cite{peng2014loss,Peng2014parity,Sahin2019PT,Arruda2020Controlling}.

\section{Results and discussions}

{\quad}Our system is composed of a whispering-gallery-mode resonator that supports periodic EPs~\cite{Peng2016chiral,chen2017exceptional}. The asymmetric coupling between the frequency-degenerate CW and CCW modes is tuned by controlling the relative size and position of two nanotips or Rayleigh scatterers placed within the mode volume of the resonator. The first nanotip induces a symmetric coupling between the CW and CCW modes and lifts their frequency-degeneracy, leading to mode-splitting. The second nanotip then breaks this symmetry, leading to periodic EPs that emerge as the relative angle between the nanotips along the boundary of the resonator is varied~\cite{Peng2016chiral} (see Figure \ref{Fig1}). This optical chiral coupling is described by the Hamiltonian:
\begin{align}
\hat{H}_\mathrm{j}=\hbar J_{12}\hat{a}_1^\dag\hat{a}_2+\hbar J_{21}\hat{a}_2^\dag\hat{a}_1,
\end{align}
where $\hat{a}_1$ and $\hat{a}_2$ are the photonic annihilation operators for the CW and CCW modes of the resonator, respectively, $J_{12(21)}=\epsilon_1+\epsilon_2e^{\pm i2\sigma\beta}$ describes the scattering rate and hence the scatterer-induced asymmetric coupling between CW and CCW modes, $2\epsilon_{j=1,2}$ is the frequency splitting induced by the $j$-th scatterer alone, and $\sigma$ and $\beta$ are the azimuthal mode number or relative angle of the scatterers.

{\quad}The Hamiltonian describing the Kerr interaction is given by~\cite{Ferretti2012Single,Vernon2015Strongly,Choi2017Self}:
\begin{align}
\hat{H}_\mathrm{k}=\sum_{j=1,2}\hbar\chi\hat{a}_j^\dag\hat{a}_j^\dag\hat{a}_j\hat{a}_j,\quad \chi=\frac{3\hbar\omega^2\chi^{(3)}}{4\varepsilon_0\varepsilon_r^2V_\mathrm{eff}},
\end{align}
where $\varepsilon_0$ ($\varepsilon_r$) is the vacuum (relative) permittivity, $\chi^{(3)}$ is the nonlinear susceptibility, and $V_\mathrm{eff}$ is the mode volume. In addition to photonic structures made from highly nonlinear materials~\cite{hales2018third,Heuck2020Controlled,Alam2016Large,zielinska2017self,Choi2017Self}, Kerr-type nonlinearity can be achieved in cavity or circuit QED systems~\cite{Birnbaum2005,Kirchmair2013Observation,gu2017microwave}, cavity free systems~\cite{Xia2018Cavity}, magnon devices~\cite{wang2018bistability,Zhang2021Exceptional}, and optomechanics~\cite{Gong2009Effective,rabl2011photon,lu2013quantum}. Then the effective Hamiltonian of the system is given by $\hat{H}_\mathrm{i}=\hbar\omega\hat{a}_1^\dag\hat{a}_1+\hbar\omega\hat{a}_2^\dag\hat{a}_2+\hat{H}_\mathrm{j}+\hat{H}_\mathrm{k}$, where $\omega=\omega_0+\epsilon_1+\epsilon_2$, and $\omega_0$ is the resonance frequency of the cavity. The eigenvalues of the system in the zero-, one-, and two-photon excitation subspaces are found as
\begin{align}
E_0=0, \quad E_1^\pm=\omega\pm\delta_1, \quad E_2^{\pm,0}=2\omega+2\chi+\delta_2^{\pm,0}, \nonumber \\
\delta_2^0=0, \quad \delta_1=\sqrt{J_{12}J_{21}}, \quad \delta_2^\pm=-\chi\pm\sqrt{\chi^2+4\delta_1^2}.
\end{align}
The corresponding eigenstates are:
\begin{align}
&\psi_0=\left|0,0\right\rangle, \nonumber \\
&\psi_1^\pm=\sqrt{J_{12}}\left|1,0\right\rangle\pm\sqrt{J_{21}}\left|0,1\right\rangle, \nonumber \\
&\psi_2^{\pm,0}=\sqrt{2}J_{12}\left|2,0\right\rangle+\delta_2^{\pm,0}\left|1,1\right\rangle+\sqrt{2}J_{21}\left|0,2\right\rangle. \label{EQ:psi}
\end{align}

\begin{figure}
  \includegraphics[width=\linewidth]{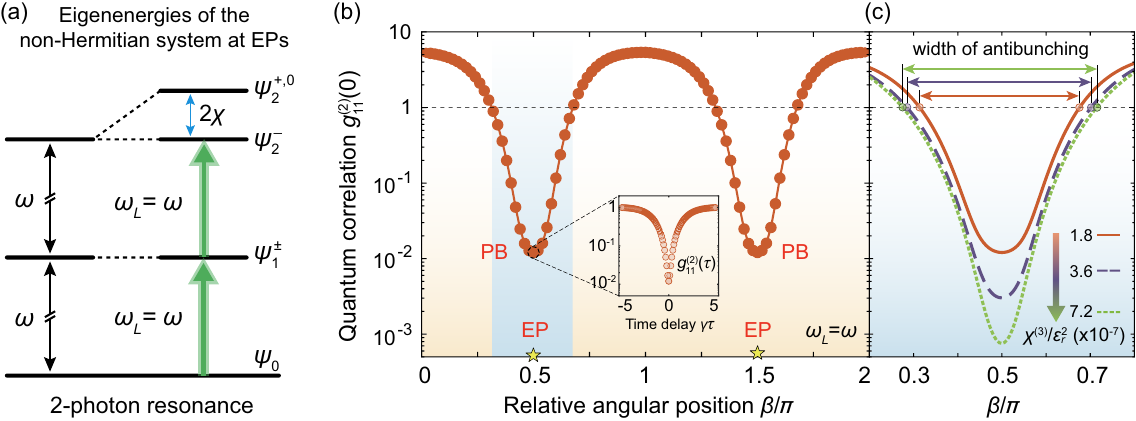}
  \caption{Non-Hermitian photon blockade (PB) at the EPs. {(a)} {The energy-level structure of the eigenstates indicates a two-photon resonance transition from $\psi_0$$\to$$\psi_2^-$ for the light with $\omega_L$$=$$\omega$.} {(b)} $g_{11}^{(2)}(0)$ of the CW mode versus $\beta$ for $\omega_{L}$$\ =\ $$\omega$. The inset shows the evolution of the second-order correlation $g^{(2)}_{11}(\tau)$ as a function of the time delay $\gamma\tau$, for $\beta/\pi$$\ =\ $$0.5$. {(c) $g^{(2)}(0)$ versus $\beta/\pi$ for different values of Kerr nonlinearity~\cite{zielinska2017self,hales2018third,Heuck2020Controlled,Alam2016Large,Choi2017Self}.} The other parameters are the same as those in Figure~\ref{Fig1}. }
  \label{Fig2}
\end{figure}

{\quad}The EPs emerge when $E_1^+=E_1^-$ and $\psi_1^+=\psi_1^-$. This imposes either $J_{21}=0$ leading to $\psi_1^\pm=\left|1,0\right\rangle$ (i.e., CW mode), or $J_{12}=0$ which leads to $\psi_1^\pm=\left|0,1\right\rangle$ (i.e., CCW mode). {For example, when $J_{12}=0$, we have
\begin{align}
&\cos(2\sigma\beta)=-\left(\text{Re}[\epsilon_{1}]\text{Re}[\epsilon_{2}]+\text{Im}[\epsilon_{1}]\text{Im}[\epsilon_{2}]\right)/\left(\text{Re}[\epsilon_{2}]^2+\text{Im}[\epsilon_{2}]^2\right), \nonumber\\
&\sin(2\sigma\beta)=-\left(\text{Re}[\epsilon_{1}]\text{Re}[\epsilon_{2}]-\text{Im}[\epsilon_{1}]\text{Im}[\epsilon_{2}]\right)/\left(\text{Re}[\epsilon_{2}]^2+\text{Im}[\epsilon_{2}]^2\right),
\end{align}
and the corresponding $J_{21}$ is
\begin{equation}
J_{21}=\epsilon_{1}-\frac{\epsilon_{2}}{\epsilon_{2}^{*}}\epsilon_{1}^{*}.
\end{equation}
Similar results can be easily obtained for the $J_{21}=0$ case.} Therefore, the Hamiltonian EPs (HEPs), which are the spectral degeneracies of the effective non-Hermitian Hamiltonian $\hat{H}_\mathrm{i}$, are periodically located at [see Figure~\ref{Fig1}(a)]
\begin{equation}
\beta_\mathrm{EP}=\frac{z\pi}{2\sigma}\pm\frac{\arg{(\epsilon_1)}-\arg{(\epsilon_2)}}{2\sigma},\ \ z=\pm1,\,\pm3,\,\dots
\end{equation}
{with ${\epsilon_1}/{\epsilon_1^*}\neq{\epsilon_2}/{\epsilon_2^*}$, and $+$ ($-$) corresponds to the case with $J_{12}=0$ and $J_{21}\neq0$ ($J_{21}=0$ and $J_{12}\neq0$). We also note that these chiral EPs can be closely related to the existence of hidden parity-time symmetry, as shown in a very recent work~\cite{Hashemi2021New}.}

\subsection{Fully quantum model}
{\quad}The non-Hermitian Hamiltonian $\hat{H}_\mathrm{i}$ does not take into account the effect of quantum jumps and the associated quantum noise. Thus, it provides only the semiclassical picture of the process. For a fully quantum picture, one should resort to the EPs of the system's Liouvillian~\cite{Minganti2019quantum}. For this purpose, we rewrite $\hat{H}_\mathrm{i}$ as the sum of a Hermitian and an anti-Hermitian part as $\hat{H}_\mathrm{i}=\hat{H}_+^\mathrm{i}+\hat{H}_-^\mathrm{i}$, with $(\hat{H}_\pm^\mathrm{i})^\dag=\pm\hat{H}_\pm^\mathrm{i}$, and use the Lindblad master-equation approach with the Liouvillian superoperator $\mathcal{L}$ given by~\cite{Minganti2019quantum}
\begin{equation}\label{soL}
\mathcal{L}\hat{\rho}=-i(\hat{H}^{\mathrm{i}}_+\hat{\rho}-\hat{\rho}\hat{H}^{\mathrm{i}}_+)+\sum_j\mathcal{D}(\hat{\rho},\hat{A}_{j})+\mathcal{D}(\hat{\rho},\hat{\Gamma}),
\end{equation}
where $\mathcal{D}(\hat{\rho},\hat{A}_{j})=\hat{A}_{j}\hat{\rho}\hat{A}_{j}^{\dagger}-\hat{A}_{j}^{\dagger}\hat{A}_{j}\hat{\rho}/2-\hat{\rho}\hat{A}_{j}^{\dagger}\hat{A}_{j}/2$
are the dissipators associated with the jump operators $\hat{A}_{j}=\sqrt{\gamma}\hat{a}_{j}$, and $\hat{\Gamma}=\sqrt{-2i\hat{H}_{-}^{\mathrm{i}}}$ is the additional jump operator. We then find the Liouvillian exceptional points (LEPs) as the degeneracies of the Liouvillian superoperator by solving the equation~\cite{Minganti2019quantum}:
\begin{equation}\label{eigenL}
\mathcal{L}\hat{\rho}_i=\lambda_i\hat{\rho}_i,
\end{equation}
where $\lambda_i$ and $\hat{\rho}_i$ are the eigenvalues and the corresponding eigenstates of $\mathcal{L}$.

{\quad}As seen in the spectra of the Liouvillian superoperator $\mathcal{L}$ and that of the effective Hamiltonian $\hat{H}_\mathrm{i}$ depicted in Figure~\ref{Fig1}(a), the positions of the LEPs and HEPs coincide, that is they occur at the same values of $\beta$. {The features of EPs can also be clearly seen in the excitation spectrum, where two spectrally separated resonance modes become overlap~\cite{chen2017exceptional,Sahin2019PT}. Moreover, under weak-driving condition, the occurrence of single-PB is closely related to the resonance mode in the excitation spectrum~\cite{rabl2011photon}.}

{\quad}We see that both HEPs and LEPs emerge at $\beta=\pi/2$ and $3\pi/2$ in the cavity excitation spectrum, see Figure \ref{Fig1}(b), with a good agreement between the semiclassical and fully quantum approaches. {This agreement means that, in our system, the role of quantum jumps can be safely ignored when locating the positions of EPs. Thus the semiclassical HEPs give a good approximation for the fully quantum LEPs in this case. This agreement was also found in e.g., a quantum non-Hermitian system with coupled bosonic modes, with the same dynamic equations derived from the Hamiltonian and the master equation~\cite{Minganti2019quantum}. Nevertheless, we stress that the role of quantum jumps should always be checked in a specific quantum EP system. We note that, e.g., in a quantum two-level system~\cite{Naghiloo2019quantum,Minganti2019quantum}, significant differences between HEPs and LEPs were revealed, which means that in that case only LEPs can be reliably considered (see more details in Ref.~\cite{Minganti2019quantum}).}

\begin{figure}[t!]
  \includegraphics[width=\linewidth]{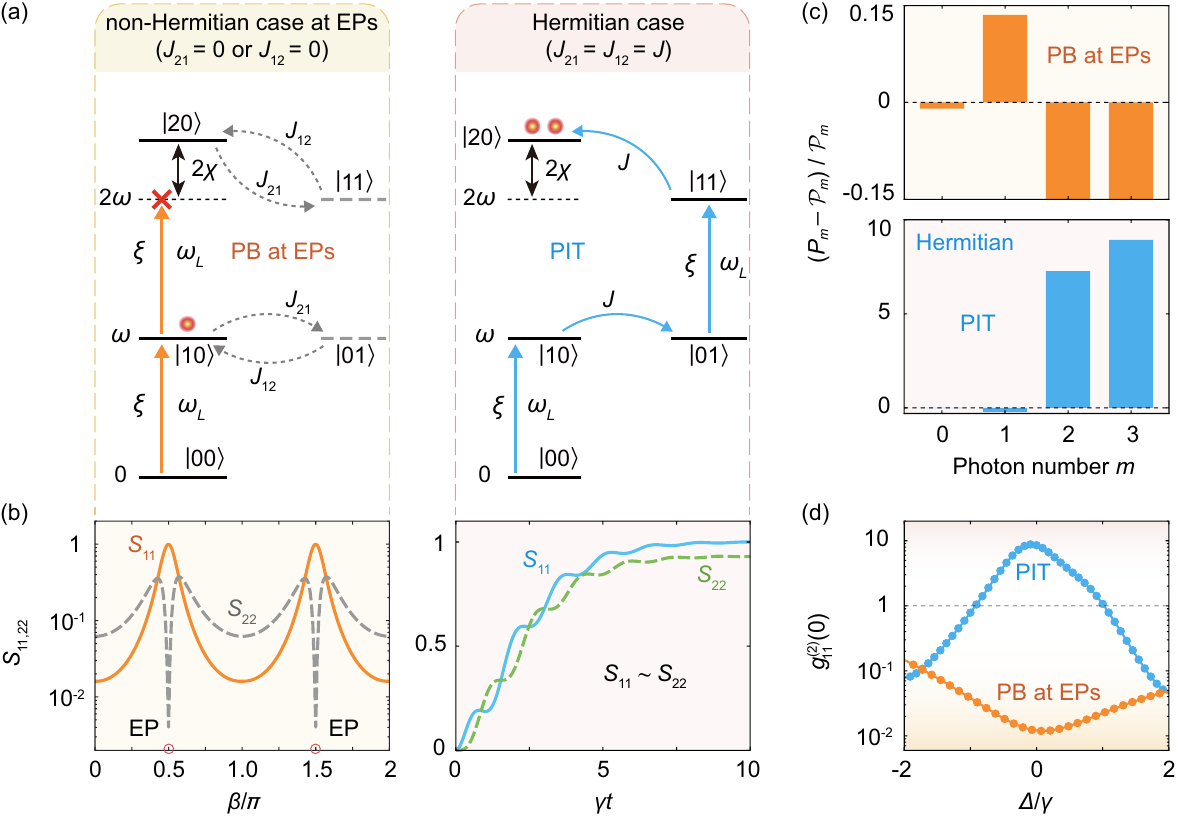}
  \caption{{(a)} The excitation pathway of the non-Hermitian case (left panel) shows the origin of {photon blockade (PB) at EPs}. The right panel shows the excitation pathway of the Hermitian system realized by one nanotip positioned near a resonator for $J_{21}$$\ =\ $$J_{12}$$\ =\ $$J$. {(b)} The normalized cavity excitation spectra of the CW ($S_{11}$) and CCW ($S_{22}$) modes. In the non-Hermitian case (left panel), $S_{11}$ (orange solid curve) and $S_{22}$ (grey dashed curve) are obtained for $\omega_L$$\ =\ $$\omega$. In the Hermitian case (right panel), $S_{11}$ (blue solid curve) and $S_{22}$ (green dashed curve) are obtained for $J/\gamma$$\ =\ $$2$. {(c)} The deviations of the photon distribution $P_m$ from the standard Poisson distribution $\mathcal{P}_m$ with the same mean photon number $m$ for $\omega_L$$\ =\ $$\omega$ ($\Delta$$\ =\ $$0$). The upper panel shows {PB at EPs} ($\beta/\pi$$\ =\ $$0.5$, orange bars), while the lower panel shows photon-induced tunneling (PIT) in the Hermitian case for $J/\gamma$$\ =\ $$2$ (blue bars). {(d)} Photon correlation $g^{(2)}_{11}(0)$ versus frequency detuning $\Delta/\gamma$ for the non-Hermitian (orange curve) and Hermitian (blue curve) cases.}
  \label{Fig3}
\end{figure}

\begin{figure}[t!]
  \includegraphics[width=\linewidth]{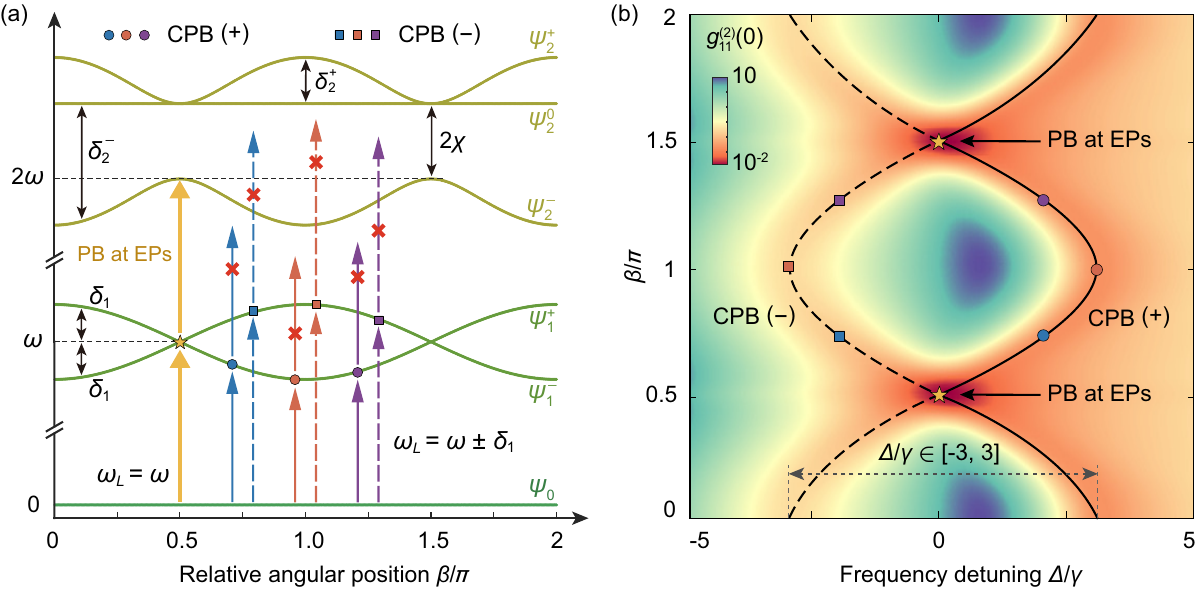}
  \caption{{(a)} Eigenenergy spectrum as a function of $\beta$. {Photon blockade (PB) at EPs} (yellow star) emerges with $\omega_L$$\ =\ $$\omega$ (two-photon resonance, yellow solid arrows). Conventional photon blockade [CPB($\pm$)] induced by the anharmonic energy-level structure with single-photon resonance occurs for $\omega_L$$\ =\ $$\omega$$\ \pm\ $$\delta_1$ for different $\beta$. The red crosses indicate suppressions of two-photon resonances. {(b)} Photon correlation $g_{11}^{(2)}(0)$ obtained as a function of $\Delta/\gamma$ and $\beta/\pi$. CPB ($\pm$) (black curves) and {PB at EPs} (yellow star) indicate single photons with frequencies in the range $[\omega-3\gamma,\ \omega+3\gamma]$. The colored circles and squares are examples for CPB ($+$) and CPB ($-$), respectively. }
  \label{Fig4}
\end{figure}

{\quad}In the frame rotating with the drive frequency $\omega_L$, the Hamiltonian of the system becomes
\begin{equation}
\hat{H}=\Delta(\hat{a}_1^\dag\hat{a}_1+\hat{a}_2^\dag\hat{a}_2)+\hat{H}_\mathrm{j}+\hat{H}_\mathrm{k}+\xi(\hat{a}_1+\hat{a}_1^\dag),
\end{equation}
where $\Delta=\omega-\omega_L$, $\xi=[{\gamma_\mathrm{ex}P_\mathrm{in}/(\hbar\omega_L)}]^{1/2}$ is the drive amplitude with laser power $P_\mathrm{in}$, and $\gamma_\mathrm{ex}$ is the resonator-waveguide coupling rate. The total cavity loss is given by $\gamma=\gamma_\mathrm{ex}+\gamma_0$, where $\gamma_0$ denotes the intrinsic losses of the resonator with the intrinsic quality factor $Q_0=\omega_0/\gamma_0$. The experimentally accessible parameters are chosen
as: $V_{\text{eff}}=150\,\mu\mathrm{m}^3$~\cite{vahala2003optical,spillane2005ultrahigh}, $Q_0=10^{10}$~\cite{pavlov2017pulse,huet2016millisecond}, $\chi^{(3)}/\varepsilon_r^2=1.8\times10^{-17}\ \mathrm{m}^2/\mathrm{V}^2$~\cite{hales2018third,Heuck2020Controlled}, $\epsilon_1/\gamma=1.5-i0.1$, $\epsilon_2/\gamma=1.485-i0.14$~\cite{Peng2016chiral}, $\lambda=1550\,\mathrm{nm}$, and $P_{\text{in}}=4\,\mathrm{fW}$~\cite{schuster2008nonlinear}. The quality factor of the ring resonator $Q_0$ has reached $10^{12}$~\cite{pavlov2017pulse,huet2016millisecond}, and $\chi^{(3)}/\varepsilon_r^2=2.0\times10^{-17}\ \mathrm{m}^2/\mathrm{V}^2$ for the semiconductor materials with GaAs~\cite{hales2018third,Heuck2020Controlled}. The materials with indium tin oxide can reach $\chi^{(3)}/\varepsilon_r^2=2.12\times10^{-17}\ \mathrm{m}^2/\mathrm{V}^2$~\cite{Alam2016Large}, and $\chi^{(3)}$ can be further enhanced to $2\times10^{-11}\ \mathrm{m}/\mathrm{V}^2$ by introducing other materials~\cite{zielinska2017self,Choi2017Self}.

{\quad}Now we study the full quantum dynamics of the system by considering the effects of quantum jumps, based on the Liouvillian superoperator $\mathcal{L}$ with $\hat{H}=\hat{H}_++\hat{H}_-$:
\begin{equation}
\dot{\hat{\rho}}=-i[\hat{H}_{+},\hat{\rho}]+\sum_{j}\mathcal{D}(\hat{\rho},\hat{a}_{j})+\mathcal{D}(\hat{\rho},\hat{\Gamma}),
\end{equation}
where $\hat{\rho}\left(t\right)$ is the normalized density matrix of the system at time $t$, with $\mathrm{tr}(\hat{\rho})=1$. We analyze the cavity excitation spectrum $S_{jj}(t)=\langle\hat{a}_{j}^{\dagger}(t)\hat{a}_{j}(t)\rangle/n_0$ ($j=1,2$) and $S_{jj}(\Delta)=\lim_{t\to\infty}S_{jj}(t)$ with $n_0=\xi^2/\gamma^2$, as well as the quantum second-order correlation $g_{11}^{(2)}(0)=\langle\hat{a}_{1}^{\dagger2}\hat{a}_{1}^2\rangle/\langle\hat{a}_{1}^{\dagger}\hat{a}_{1}\rangle^2$, which can be measured in experiments~\cite{Birnbaum2005,Faraon2008}. The condition $g_{11}^{(2)}(0)<1$ [$g_{11}^{(2)}(0)>1$] characterizes PB (photon-induced tunneling) with sub-Poissonian (super-Poissonian) photon-number statistics or photon antibunching (bunching)~\cite{Birnbaum2005,Faraon2008,Kowalewska2019Two}. Photon antibunching can also refer to two-time correlation effects~\cite{Zou90}, i.e., $g_{11}^{(2)}(0)< g_{11}^{(2)}(\tau)$.

\subsection{Photon blockade with exceptional points}

{\quad}We first explore the quantum behavior of the system at EPs emerging at $\beta=\pi/2$ or $3\pi/2$. {Figure~\ref{Fig2}(a) shows the energy-level structure of the eigenstates of the non-Hermitian system at the EPs, which indicates a two-photon resonance transition from $\psi_0\to\psi_2^-$ for the light with $\omega_L=\omega$. Here, the eigenstates $\psi_0$, $\psi_1^\pm$, and $\psi_2^{\pm,0}$ are the superposition states of the CW and CCW cavity modes [Equation~(\ref{EQ:psi})]. Very interestingly, under such a two-photon resonance condition, we find the single-PB effect, as clearly shown in Figure~\ref{Fig2}(b).} We find that $g^{(2)}(0)\sim0.012\ll1$ and $g_{11}^{(2)}(0)< g_{11}^{(2)}(\tau)$, indicating strongly antibunched single photons at EPs, which is a signature of single-PB. {Moreover, the width of the antibunching region related to this PB at EPs can be extended by enhancing Kerr nonlinearity [Figure~\ref{Fig2}(c)]. This single-PB effect with two-photon resonance, due to the interplay of the EPs and Kerr nonlinearity, can be observed in the energy-level diagram of the bare states of the system [Fig.~\ref{Fig3}(a), left panel], which is otherwise impossible since the two-photon resonance generally result in the PIT in conventional Hermitian systems [Fig.~\ref{Fig3}(a), right panel]~\cite{Faraon2008,Kowalewska2019Two}.} 

{\quad}To understand the physical mechanism behind such a counterintuitive effect, as well as the difference with the Hermitian case, we analyze the excitation pathways, as shown in Figure~\ref{Fig3}. At the EPs of the system, the CCW (CW) mode couples to the CW (CCW) mode, i.e., $J_{21}=0$ and $J_{12}\neq 0$ ($J_{12}=0$ and $J_{21}\neq 0$); resulting in a predominantly CW (CCW) propagating mode. For $J_{21}=0$, we have only $\psi_1=\left|1,0\right\rangle$ with eigenenergy $E_1=\omega$ in the one-photon subspace, as well as $\psi_2^{+,0}=\sqrt{2}J_{12}\left|2,0\right\rangle$ and $\psi_2^{-}=\sqrt{2}J_{12}\left|2,0\right\rangle-2\chi\left|1,1\right\rangle$, with eigenenergies $E_2^{+,0}=2\omega+2\chi$ and $E_2^{-}=2\omega$, respectively, in the two-photon space. Light input in the CW direction with $\omega_L=\omega$ resonantly couples to the transition $|00\rangle\to\psi_1=\left|1,0\right\rangle$. The transitions from $\psi_1=\left|1,0\right\rangle$ to $\psi_2^{+,0}=\sqrt{2}J_{12}\left|2,0\right\rangle$ are forbidden because the energies of these states are detuned by $2\chi$ from the two-photon resonance energy of $2\omega$ [see Figure~\ref{Fig3}(a)].

{\quad}One may think that a transition to the two photon state $\left|2,0\right\rangle$ is possible because the eigenenergy of $\psi_2^{-}$ coincides with the two-photon resonance energy of $2\omega$. However a closer look reveals two things: First, the transition to the $\left|2,0\right\rangle$ state is governed by $J_{12}$, which is negligible in a system with CW drive and in a predominantly CW mode [see $S_{11}\gg S_{22}\sim0$ in the left panel of Figure~\ref{Fig3}(b)]. Second, the {$\left|1,0\right\rangle$} state is intensively populated in the strongly nonlinear system ($\chi>\gamma$) under weak-driving condition ($\xi<\gamma$). Thus, once a photon is coupled into the CW mode $\left|1,0\right\rangle$, it suppresses the probability of the second photon with the same frequency going into the CW mode $\left|2,0\right\rangle$. As a result, {the interplay of EP-induced chirality and the Kerr-nonlinearity-induced anharmonic energy-level structure} produces an effective PB of the CW mode. This {PB at EPs} is confirmed by the enhancement of the single-photon state and the suppression of two- or more-photon states, which is clearly seen when the probabilities of the single-photon state ($P_1$) and more-photon states [$P_m(m>1)$] are compared with the Poisson distribution $\mathcal{P}_m$, as shown in Figure~\ref{Fig3}(c).

{\quad}In a system with one nanotip, the optical coupling is Hermitian and symmetric, $J_{21}=J_{12}=J$, $S_{11}\sim S_{22}$ [see the right panel of Figure~\ref{Fig3}(b)]. The CW input light with $\omega_L=\omega$ leads to one-photon excitation in the CW and CCW modes. For the input coupled to the CW mode, $\left|0,0\right\rangle\stackrel{\xi}{\longrightarrow}\left|1,0\right\rangle$, the coupling $J$ enables the transition of the single photon from the CW mode to the CCW mode as $\left|1,0\right\rangle\stackrel{J}{\longrightarrow}\left|0,1\right\rangle$. Subsequently, a second photon can couple to the CW mode through $\xi$ leading to $\left|1,1\right\rangle$. Finally, the photon in the CCW mode completes the transition back to the CW mode through the action of $J$, which results in the state $\left|2,0\right\rangle$. Thus, in this case, two photons can be absorbed in the CW mode through the pathway [Figure~\ref{Fig3}(a)]:
\begin{equation}
\left|0,0\right\rangle\stackrel{\xi}{\longrightarrow}\left|1,0\right\rangle\stackrel{J}{\longrightarrow}\left|0,1\right\rangle\stackrel{\xi}{\longrightarrow}\left|1,1\right\rangle\stackrel{\sqrt{2}J}{\longrightarrow}\left|2,0\right\rangle.
\end{equation}
As a result, two- or more-photon probabilities $P_m(m\geq2)$ are enhanced leading to photon-induced tunneling [Figure~\ref{Fig3}(c)]. We conclude that for light with $\omega_L=\omega$, \emph{{PB} with strong antibunched single photons emerges at EPs, while a bunched stream occurs in the Hermitian case} [Figure~\ref{Fig3}(d)]. This behavior is also seen in Figure~\ref{Fig2}(b). By tuning the system close to or away from EPs using $\beta$ as a knob, one can vary the value of $g^{(2)}_{11}(0)$ from bunching with $g^{(2)}_{11}(0)\sim5.37$ to anti-bunching light with $g^{(2)}_{11}(0)\sim0.012$, i.e., up to \emph{3 orders of magnitude}.

{\quad}Finally, we study the case where the drive input light has the frequency $\omega_L=\omega\pm\delta_1$. As seen in Figure~\ref{Fig4}, one can observe \emph{conventional photon blockade} [CPB($\pm$)] at various values of the relative angular position $\beta$ between the scatterers. The origin of CPB($\pm$) can be understood from the anharmonic energy-level structure induced by the strong nonlinearity [see Figure~\ref{Fig4}(a)]. Input lights with frequencies $\omega_L=\omega\pm\delta_1$ are resonantly coupled to the transitions from $\psi_0$ to $\psi_1^\pm$; however, the transitions from the single-photon subspace $\psi_1^\pm$ to the two-photon subspace $\psi_2^{\pm,0}$ are forbidden because the energies of these transitions are largely detuned from the two-photon resonance energy $2\omega_L$. As a result, the system can absorb only one photon, leading to PB and antibunched single photons with $g^{(2)}_{11}(0)<1$ [Figure~\ref{Fig4}(b)]. In contrast to the CPB with fixed frequencies in Hermitian systems~\cite{Birnbaum2005,Faraon2008,Lang2011,Hoffman2011,rabl2011photon}, the CPB ($\pm$) and {PB at EPs} in our system can generate single photons with frequencies in the range $[\omega-3\gamma,\omega+3\gamma]$. The frequency of these photons can be tuned by varying $\beta$. This functionality can be useful in constructing frequency-tunable single-photon devices.

\section{Conclusions and outlook}

{\quad}In conclusion, we have shown that EPs provide a powerful new tool for quantum engineering of single photons in nonlinear systems. Our study reveals:

{\quad}(i) The interplay of EPs and Kerr nonlinearity creates a new type of single-photon blockade effect which takes place at a two-photon resonance. In stark contrast with Hermitian two-photon resonances which exhibit bunched light, mode-coalescence at EPs,  {together with the nonlinearity}, results in strongly anti-bunched single photons.

{\quad}(ii) By tuning the system towards or away from EPs, quantum correlations of photons can be well tuned from antibunching-to-bunching regimes or vice versa. We note that, together with other methods (e.g., tuning the optical phases, see Ref.~\cite{Casalengua2020Conventional} for details), more degrees of freedom can be achieved in steering quantum behaviors of single photons.

{\quad}(iii) The frequency of single photons can be tuned in the range $[\omega-3\gamma,\omega+3\gamma]$ by tuning the relative angle of the scatterers $\beta$ and the optical detuning $\Delta$. 

{\quad}These observations suggest the important role of EPs in tuning quantum effects of various systems and achieving unconventional devices such as EP-controlled single-photon sources or EP-enhanced quantum sensors. {We also note that the two-photon resonance antibunching, as revealed here, is also possible to occur in a wide range of other chiral quantum systems, such as photonic-crystal membranes and resonator-emitter systems~\cite{Lodahl2017Chiral}.} In a broader view, our work can stimulate more explorations on EP-engineered quantum or topological effects~\cite{Zhong2019Sensing,Hafezi2011robust,Hafezi2013Imaging}. With quantum and/or topological EP devices at hand to manipulate strongly correlated photons, one may envision exciting possibilities for building unconventional quantum information architectures.

\medskip
\textbf{Acknowledgements} \par 
The authors thank Wei Qin, Fen Zou, Yunlan Zuo and Adam Miranowicz for helpful discussions. {\c{S}}.K.{\"O}. is supported by the Air Force Office of Scientific Research (AFOSR) Multidisciplinary University Research Initiative (MURI) grant (Award No. FA9550-21-1-0202). H.J. is supported by the NSFC (11935006 and 11774086). R.H. is supported by the Japan Society for the Promotion of Science (JSPS) Postdoctoral Fellowships for Research in Japan (No. P22018). J.-Q.L. is supported by NSFC (11774087 and 11822501). L.-M.K. is supported by NSFC (1217050862 and 11434011). F.N. is supported by Nippon Telegraph and Telephone Corporation (NTT) Research, the Japan Science and Technology Agency (JST) [via the Quantum Leap Flagship Program (Q-LEAP), and the Moonshot R\&D Grant Number JPMJMS2061], the Japan Society for the Promotion of Science (JSPS) [via the Grants-in-Aid for Scientific Research (KAKENHI) Grant No. JP20H00134], the Army Research Office (ARO) (Grant No. W911NF-18-1-0358), the Asian Office of Aerospace Research and Development (AOARD) (via Grant No. FA2386-20-1-4069), and the Foundational Questions Institute Fund (FQXi) via Grant No. FQXi-IAF19-06.


%
\bibliographystyle{ieeetr}

\end{document}